\documentclass[twocolumn]{article}

\usepackage{amsmath,amssymb,url}
\usepackage{graphicx}
\usepackage[numbers]{natbib}

\newcommand{\mm}{{\boldsymbol{m}}}
\newcommand{\uu}{{\boldsymbol{u}}}
\newcommand{\xx}{{\boldsymbol{x}}}
\newcommand{\yy}{{\boldsymbol{y}}}
\newcommand{\xixi}{{\boldsymbol{\xi}}}
\DeclareMathOperator{\conv}{conv}
\newcommand{\abs}[1]{\left\lvert{#1}\right\rvert}
\providecommand{\od}[3][{}]%
{\mathchoice%
  {\frac{d^{#1}{#2}}{d{#3}^{#1}}}
  {d^{#1}{#2}/d{#3}^{#1}}
  {d^{#1}{#2}/d{#3}^{#1}}
  {d^{#1}{#2}/d{#3}^{#1}}} 
\providecommand{\pd}[3][{}]%
{\mathchoice%
  {\frac{\partial^{#1}{#2}}{\partial{#3}^{#1}}}
  {\partial^{#1}{#2}/\partial{#3}^{#1}}
  {\partial^{#1}{#2}/\partial{#3}^{#1}}
  {\partial^{#1}{#2}/\partial{#3}^{#1}}} 

\defcitealias{S86}{S}
\defcitealias{ERS96}{ERS}

\title{Ballistic aggregation in symmetric and non-symmetric flows}

\author{A.~A.~Andrievski{\u\i}${}^1$
  \and S.~N.~Gurbatov${}^{2,3}$
  \and A.~N.~Sobolevski{\u\i}${}^{1,3,4}$%
  \thanks{Corresponding author: \protect\url{ansobol@obs-nice.fr}}
}

\date{}

\begin{document}

\maketitle

\footnotetext[1]{Physics Department, M.~V.~Lomonosov Moscow State
  University, Russia}
\footnotetext[2]{Radiophysics Department, N.~N.~Lobachevski{\u\i}
  Nizhny Novgorod State University, Russia}
\footnotetext[3]{Observatoire de la C{\^o}te d'Azur, Nice, France.}
\footnotetext[4]{International Institute for Earthquake Prediction
  Theory and Mathematical Geophysics of the Russian Academy of
  Sciences, Moscow, Russia}

\begin{abstract}
  Explicit solutions for ballistic aggregation of dust-like matter,
  whose particles stick inelastically upon collisions, are considered.
  This system provides a model of large-scale structure formation in
  cosmology within the Zel'dovich approximation.  In particular we
  show the equivalence of two different representations of solutions
  proposed in \citep{S86,ERS96} for a flat 1D flow, extend these
  representations to cylindrically or spherically symmetric flows, and
  provide explicit counterexamples showing how exactly these
  representations break down in the case of non-symmetric flow.
\end{abstract}

\section{Introduction}
\label{sec:intro}

We consider here explicit solutions for ballistic aggregation of
dust-like matter, with particles that stick upon collisions absolutely
inelastically.  This system may be considered as a model for the large
structure formation in the Universe within the Zel'dovich
approximation~\citep{Z70}.  First we briefly recall how this model
comes about.

Consider a flat Einstein--de Sitter universe containing massive
dust-like matter that moves in a self-consistent gravitational field.
This matter participates in two types of motion, the homogeneous
cosmological Hubble expansion and the evolution of local perturbations
of density and velocity, which leads to formation of large-scale
inhomogeneities.

To describe the latter kind of dynamics in coordinates comoving with the
Hubble expansion, Ya.~B.~Zel'dovich proposed~\citep{Z70} to use a
nonlinear function of time, the perturbation growth factor, instead of
the cosmological time.  In the new time, the gravitational force turns out
to be approximately balanced by the `Hubble drag' caused by the
continuous expansion of the spatial scale.  Within this approximation
fluid elements move ballistically, and the evolution of local
perturbations is described by the following system of equations:
\begin{gather}
  \label{eq:mass}
  \pd\rho t + \nabla\cdot(\rho\uu) = 0,\\
  \label{eq:momentum}
  \pd{(\rho\uu)}t + \nabla\cdot(\rho\uu\otimes\uu) = 0,
\end{gather}
expressing conservation of respectively mass and momentum.  Here \(t\)
is the new time parameter, \(\rho(\xx, t)\) is the mass density field,
\(\uu(\xx, t)\) is the velocity field, \(\xx\) is the spatial position
comoving with the Hubble expansion, and \(\uu\otimes\uu\) denotes the
tensor with components~\(u_iu_j\).

\begin{figure}
  \centering\includegraphics[width=6cm]{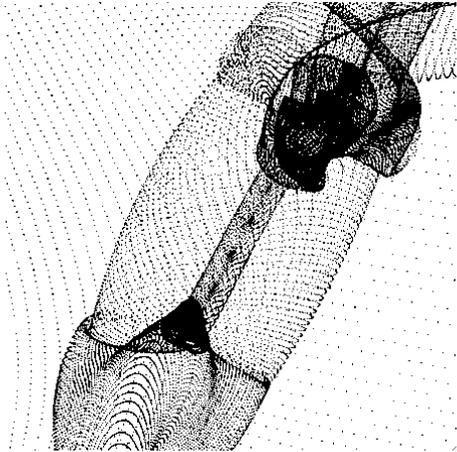}
  \caption{Matter concentration in regions of multi-streamed flow
    (2D numerical simulations of~\citep{MS89}).}
  \label{fig:aggreg}
\end{figure}

When trajectories of fluid elements cross, the velocity field ceases
to be single-valued, and several streams passing through the same
spatial location form.  Numerical simulations show (see e.g.\
\citep{MS89,WG90}) that the gravitational interaction between streams
confines the matter to domains of relatively small width and high
density (Fig.~\ref{fig:aggreg} on page~\pageref{fig:aggreg}), causing
it to aggregate in wall-like, filament-like, and cluster-like
structures.  This process is called \emph{ballistic aggregation}.

In \citep{GS84,GSS89} (see also the monograph \cite{GMS92} and
\cite{GM03}) the Burgers equation
\begin{equation*}
  \pd\uu t + (\uu\cdot\nabla)\uu = \mu\Delta\uu
\end{equation*}
was proposed to describe ballistic aggregation.  Indeed, in the
vanishing viscosity limit \(\mu\to 0\) solutions to the Burgers
equation develop singularities in the form of infinitesimally thin
walls, filaments, and point clusters, whose shape is tantalizingly
similar to the large-scale cosmological structure.  However this
agreement is limited, as the Burgers equation is written in terms of
the velocity field and fails to account for conservation of mass or
momentum.  Quantitatively, direct \(N\)-body simulations of
self-gravitating matter disagree with solutions of the Burgers
equation for identical initial data.

These limitations can be overcome in the case of 1D flow with flat
symmetry (note that in the 1D case a thorough analytical study is
possible even without the aggregation approximation
\cite{AFGM03}). Several authors \citep{S86,MP94,ERS96,BG98}
independently introduced exact solutions for ballistic aggregation in
a 1D flow of particles that stick upon collisions with exact
conservation of mass and momentum.  In particular, in \citep{S86}
and~\citep{ERS96} these solutions were expressed in the form of
variational principles, so that the density and velocity fields at any
given time could be constructed by minimization of certain
functionals.

It is the purpose of the present work to explore the possibility of
extending these results to the case where there is no flat symmetry.
In Section~\ref{sec:flat} we prove the equivalence of the variational
principles proposed in~\citepalias{ERS96}
and~\citepalias{S86}\footnote{In the sequel, \citep{S86}
  and~\citep{ERS96} are referred to as \citepalias{S86} and
  \citepalias{ERS96} correspondingly.} in the 1D case with flat symmetry.  To make the
exposition self-contained we recall the necessary details of
constructions of \citepalias{S86} and~\citepalias{ERS96}, in
particular because Ref.~\citepalias{S86} is rather difficult to
obtain.  In Section~\ref{sec:spherical} the variational principles are
extended to the case of cylindrical and spherical symmetry.

Both variational principles \citepalias{S86} and~\citepalias{ERS96}
can be extended formally to the case of 2D or 3D non-symmetric flow,
but the corresponding constructions no longer coincide.  Moreover,
there exist counterexamples showing that neither variational principle
is valid in its extended version
already for non-symmetric 2D flows.  The multidimensional
extensions of the variational principles \citepalias{S86}
and~\citepalias{ERS96} and explicit counterexamples are presented in
Section~\ref{sec:counterexamples}.

The second and third authors are grateful to the French Ministry of
education for its support.  The work of SG was also supported by RFBR
(grant 05--02--16517) and the Russian Leading Schools in Science support
program (NSh-5200.2006.2).  The work of AS was supported by RFBR
(grant 05--01--00824).

\section{Ballistic aggregation in 1D}
\label{sec:flat}

\subsection{Clusters and free particles}
\label{sec:clusters}

In the case of flat symmetry the density field~\(\rho(x, t)\) and the
velocity field~\(u(x, t)\) both depend on a single coordinate.  Mass
concentration due to inelastic collisions results in the formation of
parallel massive `walls' in three-dimensional space, which can be
considered as point-like structures (clusters) in the \(x\) axis.
We consider first the evolution of a purely discrete mass distribution.

Let \(N\) particles in the \(x\) axis be located initially (i.e., at
\(t = 0\)) at points \(y_1 < y_2 < \dots < y_N\) and have velocities
\(u_1, u_2, \dots, u_N\) and masses \(m_1, m_2, \dots, m_N\).  As time
passes, \(i\)-th particle describes the trajectory \(x_i(t) = y_i + u_i
t\) unless it collides with another particle. After an inelastic
collision there forms a cluster containing the whole mass of the group of
particles that stuck together.  Since trajectories of free particles
cannot cross, such clusters will always contain groups of particles
numbered contiguously, i.e., with \(i^- \le i \le i^+\).  Mass,
velocity and coordinate of each cluster coincide with the total mass
and center-of-mass velocity and coordinate of the corresponding group
of particles, defined for the free motion:
\begin{equation}
  \label{eq:clusterdiscrete}
  \begin{gathered}
    m = \sum_{i^- \le i \le i^+} m_i, \quad
    u = \frac 1m \sum_{i^- \le i \le i^+} u_im_i,\\
    x = \frac 1m \sum_{i^- \le i \le i^+} (y_i + u_i t)m_i.
  \end{gathered}
\end{equation}

In order for \(i_0\)-th particle to stay free at time~\(t > 0\), its trajectory
must not cross center-of-mass trajectories of any groups of particles
immediately adjacent to it.  Therefore for any \(i', i''\)
such that \(1 \le i' < i_0 < i'' \le N\), the coordinate
\(x_{i_0}(t)\) must satisfy the inequalities
\begin{equation}
  \label{eq:CMdisc}
  \begin{split}
    \frac 1{m'}\!\! \sum\limits_{i' \le i < i_0} (y_i + u_i t)m_i 
    &< x_{i_0}(t) = y_{i_0} + u_{i_0}t \\
    &< \frac 1{m''}\!\! \sum\limits_{i_0 < i \le i''} (y_i + u_i t)m_i,
  \end{split}
\end{equation}
where
\[
m' = \sum\limits_{i' \le i < i_0} m_i,\quad
m'' = \sum\limits_{i_0 < i \le i''} m_i.
\]
Using these conditions and equations~\eqref{eq:clusterdiscrete}, one
can find coordinates and masses of all particles and clusters at~\(t >
0\) without explicitly computing dynamics of the system at
intermediate times.  Thus equations \eqref{eq:CMdisc}
and~\eqref{eq:clusterdiscrete} give a complete solution for a discrete
initial mass distribution.

In the case of a continuous mass distribution the initial density and
velocity fields are given by the functions \(\rho_0(x)\)
and~\(u_0(x)\).  In this case one replaces summation in equations
\eqref{eq:clusterdiscrete} and~\eqref{eq:CMdisc} with integration, the
index~\(i\) with the continuous initial coordinate~\(y\), and mass of
\(i\)-th particle with the mass element \(\rho_0(y)\, dy\).
Equation~\eqref{eq:CMdisc} then assumes the form
\begin{multline}
  \label{eq:GPrinciple}
  \frac{\int_{y'}^{y_0} (y + u_0(y) t)\rho_0(y)\, dy}
  {\int_{y'}^{y_0} \rho_0(y)\, dy} \le y_0 + u_0(y_0)t\\
  \le \frac{\int_{y_0}^{y''}\!\! (y + u_0(y) t)\rho_0(y)\, dy}
  {\int_{y_0}^{y''}\rho_0(y)\, dy}.
\end{multline}
If these inequalities are satisfied for any \(y' < y_0 < y''\) and at
least one of them is strict, the particle located initially at~\(y_0\)
will stay free at \(t > 0\) and will be located at \(x(y_0, t) = y_0 +
u_0(y_0)t\).

On the other hand, let a group of particles located initially at \(y^-
< y < y^+\) stick together at time~\(t\) into a cluster surrounded by
free particles. Then mass, velocity, and coordinate of this cluster
are given by
\begin{equation}
  \label{eq:cluster}
  \begin{gathered}
    m = \int_{y^-}^{y^+}\!\!\!\!\!\! \rho_0(y)\, dy,\quad
    u = \frac 1m \int_{y^-}^{y^+}\!\!\!\!\!\! u_0(y)\rho_0(y)\, dy,\\
    x = \frac 1m \int_{y^-}^{y^+}\!\!\!\!\!\! (y + u_0(y)t)\rho_0(y)\, dy.
  \end{gathered}
\end{equation}
For any particle specified by the initial position~\(y\), equations
\eqref{eq:GPrinciple} and~\eqref{eq:cluster} together give its
position \(x(y, t)\) at time \(t>0\) without invoking the dynamics at
intermediate times.

There are two ways to construct an explicit formula for~\(x(y, t)\)
from \eqref{eq:GPrinciple} and~\eqref{eq:cluster}, which are discussed
in the following two subsections.  Both of them involve minimizing
suitable functionals and can therefore be considered as (generalized)
variational principles.

\subsection{The variational principle \citepalias{ERS96}}
\label{sec:sinai}

Introduce in \eqref{eq:GPrinciple} and~\eqref{eq:cluster} a `mass
coordinate'
\begin{equation}
  \label{eq:10}
  m(y)=\int^{y}\rho_0(\eta)\, d\eta
\end{equation}
and let
\begin{equation}
  \label{eq:11}
  \begin{gathered}
    \Phi_0(m) = \int^{y(m)}\!\!\!\!\!\!\!\!\! \eta\rho_0(\eta)\, d\eta,\
    U_0(m) = \int^{y(m)}\!\!\!\!\!\!\!\!\! u_0(\eta)\rho_0(\eta)\, d\eta,\\
    \Phi_t(m)=\Phi_0(m)+tU_0(m).
  \end{gathered}
\end{equation}
The lower limits of integration in~\eqref{eq:10}, \eqref{eq:11} may be
chosen arbitrarily.  It can easily be checked that the inverse
function~\(y(m)\) and \(u_0(y)\) have the form
\begin{equation}
  \label{eq:12}
  y(m) = \od{\Phi_0(m)}m,\quad u_0(y(m))=\od{U_0(m)}m.
\end{equation}
Hence, as long as some particle is not absorbed into a cluster, its
coordinate is
\begin{equation}
  \label{eq:3}
  y + u_0(y)t 
  = \od{\Phi_0}m + \od{(tU_0)}m = \od{\Phi_t}m.
\end{equation}
However after crossing of trajectories this representation of \(x(y,
t)\) breaks down.  To give an expression for \(x(y, t)\) that is valid
for all times, it is convenient to consider the cases of a free
particle and a cluster separately.

\begin{figure}
  \label{fig:convex}
  \centering\includegraphics{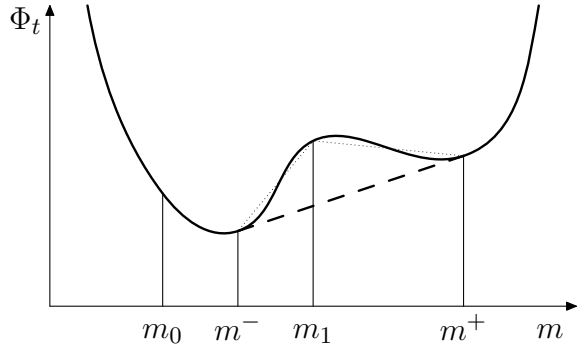}
  \caption{Construction of a solution according to the variational
    principle~\citepalias{ERS96}.  At \(m_0\)
    condition~\protect\eqref{eq:GP1} is satisfied with any
    \(m'<m_0<m''\), while at \(m_1\) it is violated (e.g.\ for \(m' =
    m^-\), \(m''=m^+\) the slope of the left dotted line is positive
    while that of the right one is negative).  On the segment~\((m^-,
    m^+)\), the convex hull~\(\conv\Phi_t(m)\) (dashed line) does not
    coincide with~\(\Phi_t(m)\).  Thus particles corresponding to
    points of the segment~\((m^-, m^+)\) stick together and form a
    cluster of the mass~\(m^+ - m^-\) by the time~\(t\).}
\end{figure}

In the new variables, condition (\ref{eq:GPrinciple}) determining if a
particle stays free at time~\(t > 0\) takes the following form: if
\(m_0 = m(y_0)\), then for any \(m' < m_0 < m''\)
\begin{equation}
  \label{eq:GP1}
  \frac {\Phi_t(m_0)-\Phi_t(m')} {m_0-m'}
  < \frac {\Phi_t(m'')-\Phi_t(m_0)} {m''-m_0}.
\end{equation}
In other words, at values of~\(m\) corresponding to free particles the
function \(\Phi_t(m)\) coincides with its convex hull \(\conv\Phi_t\),
i.e., the maximal convex function not exceeding~\(\Phi_t\)
(Fig.~\ref{fig:convex}).  One can visualize the graph of a convex hull
as the shape of an elastic thread tightly wrapped around the graph of
\(\Phi_t\) from below.  Since the function \(\Phi_t\) is
differentiable, its convex hull is also differentiable, and at common
points of their graphs the derivatives \(\od{\Phi_t}m\) and
\(\od{(\conv\Phi_t)}m\) coincide.  Therefore at these points
equation~\eqref{eq:3} takes the form
\begin{equation*}
  x(y, t) = \od{\Phi_t}m = \od{(\conv\Phi_t)}m.
\end{equation*}

Conversely, if by the time~\(t\) a group of particles initially
located in the segment \(y^- < y < y^+\) forms a cluster surrounded
with free particles, then condition~\eqref{eq:GP1} is violated at all
internal points of the corresponding segment \(m^- = m(y^-) < m < m^+
= m(y^+)\) and the graph of~\(\Phi_t(m)\) lies above the chord
connecting \(\Phi_t(m^-)\) and~\(\Phi_t(m^+)\), and thus above the
convex hull (Fig.~\ref{fig:convex}).  Moreover, \eqref{eq:cluster}
implies that
\begin{equation*}
  x = \frac{\Phi_t(m^+) - \Phi_t(m^-)}{m^+ - m^-}
  = \od{\bigl(\conv\Phi_t\bigr)}m.
\end{equation*}

Thus in clusters as well as in intervals of continuous mass
disribution the map \(x(y, t)\) is determined by the derivative of the
convex hull of~\(\Phi_t\):
\begin{equation}
  \label{eq:ERSvariational}
  x(y(m), t) = \od{\bigl(\conv\Phi_t(m)\bigr)}m.
\end{equation}
In~\citepalias{ERS96}, this expression is called the Generalized
Variational Principle by analogy with the Hopf--Lax--Ole{\u\i}nik
variational principle used to construct solutions to the Burgers
equation.  The latter principle itself is a variant of the least
action principle in mechanics.  In the present paper,
\eqref{eq:ERSvariational} will be referred to as \emph{the variational
  principle~\citepalias{ERS96}}.  Analogous representations for \(x(y,
t)\) were introduced independently in \citep{MP94} and~\citep{BG98}.

In the case of a constant initial density, when \(\rho_0(y) \equiv 1\)
and \(m \equiv y\), the variational principle \citepalias{ERS96}
coincides with the Hopf--Lax--Ole{\u\i}nik variational principle.

\subsection{The variational principle \citepalias{S86}}
\label{sec:shnirelman}

Another representation for \(x(y, t)\) was introduced a decade earlier
by A.~I.~Shnirel'man in~\citepalias{S86}.  To simplify notation,
introduce the \emph{displacement field}
\begin{equation*}
  \xi_t(y) = x(y, t) - y.
\end{equation*}
As long as trajectores of particles do not cross, \(\xi_t(y) =
u_0(y)t\) holds and the coordinate~\(x(y, t) = y + \xi_t(y)\) is
monotonic as a function of~\(y\):
\begin{equation}
  \label{eq:4}
  y' + \xi_t(y') \le y'' + \xi_t(y'')
\end{equation}
whenever \(y' < y''\).

We consider below only displacement fields for which the integral
\(\int \xi_t^2(y)\rho_0(y)\, dy\) exists.  This is the case if the
initial density~\(\rho_0(y)\) and velocity~\(u_0(y)\) (and therefore
the displacement field) either decrease at infinity fast enough or are
periodical.

Call a displacement field \emph{feasible} if trajectories of particles
do not cross, i.e., if condition~\eqref{eq:4} is satisfied for any
\(y' < y''\). According to \emph{the variational
  principle~\citepalias{S86}}, the solution \(x(y, t)\) is given by
\begin{equation*}
  x(y, t) = y + \bar\xi_t(y),
\end{equation*}
where \(\bar\xi_t(y)\) is a feasible displacement field that minimizes
the following norm of discrepancy with respect to \(u_0(y)t\):
\begin{equation}
  \label{eq:6}
  \int \abs{\bar\xi_t(y) - u_0(y)t}^2\rho_0(y)\, dy = \min.
\end{equation}

In the simplest case when trajectories of particles do not cross,
\(\bar\xi_t(y) = u_0(y)t\); this displacement field is feasible and
therefore the discrepancy vanishes.  After crossing of trajectories a
feasible displacement field minimizing~\eqref{eq:6} can no longer
coincide with~\(u_0t\).

Mathematically, minimization problem~\eqref{eq:6} is equivalent to
the orthogonal projection of the field \(u_0(y)t\) to the set of
feasible displacements, which we denoted hereafter by~\(\mathcal{X}\).
Here orthogonality is understood in the sense of the functional
scalar product
\begin{equation*}
  \xi\cdot\eta = \int \xi(y)\eta(y)\, \rho_0(y)\, dy,
\end{equation*}
and the square of the norm of discrepancy~\eqref{eq:6} is its scalar
poduct with itself:
\begin{multline}
  \label{eq:1}
  {\lVert\xi_t - u_0t\rVert}^2
  = \int \abs{\xi_t(y) -tu_0(y)}^2\rho_0(y)\, dy\\
  = (\xi_t - u_0t)\cdot (\xi_t - u_0t).
\end{multline}
Observe that feasible displacements form a convex set: if displacement
fields \(\xi_1(y)\), \(\xi_2(y)\) are feasible, then so is their
`mixture' \((1 - \alpha)\xi_1(y) + \alpha\xi_2(y)\) for any
\(0<\alpha<1\).  Moreover, a set of feasible displacements is closed:
a limit of any sequence of feasible displacements is also feasible.
It is a well-known fact of functional analysis (see, e.g.,
\citep{B71}) that an orthogonal projection to a closed
set~\(\mathcal{X}\) exists and is defined uniquely.  Therefore the
variational principle~\citepalias{S86} always gives a uniquely defined
result.

\begin{figure}
  \centering\includegraphics{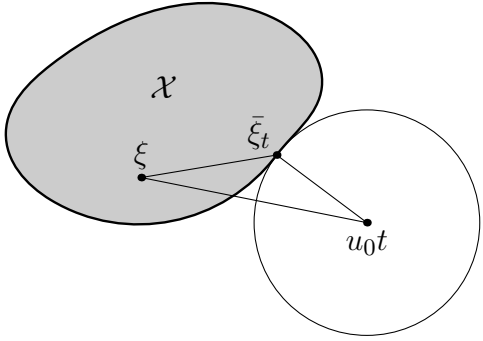}
  \caption{Orthogonal projection to a convex set~\(\mathcal{X}\) in a
    Hilbert space.  Shown is a section of the set~\(\mathcal{X}\) with
    a hyperplane passing through the elements \(u_0t\), \(\bar\xi_t\)
    and~\(\xi\).  Since the point~\(\bar\xi_t\) is closest to \(u_0t\)
    in the set~\(\mathcal{X}\), this set lies outside the sphere
    determined by the radius \((u_0t, \bar\xi_t)\).  Therefore in the
    triangle connecting \(u_0t\), \(\bar\xi_t\), and~\(\xi\) the angle
    at \(\bar\xi_t\) must be obtuse, i.e., the square of the corresponding
    segment is larger than the sum of squares of two other segments
    (inequality~\protect\eqref{eq:88b}).}
  \label{fig:orthoconvex}
\end{figure}

We now show that this result coincides with the result given by the
variational principle~\citepalias{ERS96}.  Denote the displacement
field defined by the map~\(x(y, t)\) from~\eqref{eq:ERSvariational} by
\begin{equation*}
  \tilde\xi_t(y(m)) = \od{\bigl(\conv\Phi_t(m)\bigr)}m - y(m).
\end{equation*}
It suffices to check if this displacement field
minimizes~\eqref{eq:6}, i.e., if for any other feasible
displacement~\(\xi\) in~\(\mathcal{X}\)
\begin{equation*}
  {\bigl\lVert\xi - u_0t\bigr\rVert}^2
  > {\bigl\lVert\tilde\xi_t - u_0t\bigr\rVert}^2.
\end{equation*}
It is clear from Fig.~\ref{fig:orthoconvex} that the orthogonal
projection to the closed set~\(\mathcal{X}\) must satisfy an even
stronger inequality,
\begin{equation}
  \label{eq:88b}
  {\bigl\lVert\xi - u_0t\bigr\rVert}^2
  > {\bigl\lVert\tilde\xi_t - u_0t\bigr\rVert}^2
  + {\bigl\lVert\xi - \tilde\xi_t\bigr\rVert}^2,
\end{equation}
and it is this condition that we will check.

Using~\eqref{eq:1} rewrite inequality~\eqref{eq:88b} in the form
\begin{equation}
  \label{eq:criterion}
  (\tilde\xi_t - \xi)\cdot(u_0t - \tilde\xi_t) > 0
\end{equation}
and compute its left-hand side:
\begin{equation}
  \label{eq:77}
  \begin{split}
    & (\tilde\xi_t - \xi)\cdot (u_0t - \tilde\xi_t)\\
    &\quad = \int \bigl(\tilde\xi_t(y) - \xi(y)\bigr)
    \bigl(u_0(y)t - \tilde\xi_t(y)\bigr)\rho_0(y)\, dy.
  \end{split}
\end{equation}
Observe that
\begin{equation}
  u_0t - \tilde\xi_t = y + u_0t - x(y, t) 
  = \pd{\Phi_t}m - \pd{(\conv\Phi_t)}m,
\end{equation}
where \(y = y(m)\).  This expression is nonzero on a set of
segments~\((m^-_i, m^+_i)\) of the \(m\) axis where \(\Phi_t(m) >
\conv\Phi_t(m)\) is satisfied. Therefore \eqref{eq:77} decomposes into
a sum of integrals over segments \(\bigl(y(m^-_i), y(m^+_i)\bigr) =
(y_i^-, y_i^+)\), on each of which \(x(y, t)\) is constant,
\[
x(y(m), t) = \frac{\Phi_t(m^+_i) - \Phi_t(m^-_i)}{m^+_i - m^-_i}
= x_i,
\]
and the integrand in~\eqref{eq:77} is equal to the product of the
continuous function \(\bigl(u_0(y)t - \tilde\xi_t(y)\bigr)\rho_0(y)\)
with the function
\begin{equation*}
  \tilde\xi_t(y) - \xi(y) = x_i - (y + \xi(y)),
\end{equation*}
which decreases because \(\xi(y)\) is a feasible displacement field.
According to Bonnet's formula for the average value of the product of
a monotonic and a continuous function\footnote{If the
  function~\(f(x)\) is monotonic and~\(g(x)\) is continuous, then
  there exists a point~\(c\) in the segment \((a, b)\) such that
  \(\int_a^b\! f(x)g(x)\, dx = f(a)\int_a^c\! g(x)\, dx +
  f(b)\int_c^b\!  g(x)\, dx.\)}, there exists a point~\(y_i\) in the
segment~\((y_i^-, y_i^+)\) such that
\begin{multline}
  \int_{y_i^-}^{y_i^+} \bigl(x_i - y - \xi(y)\bigr)
  \bigl(u_0(y)t - \tilde\xi_t(y)\bigr)\rho_0(y)\, dy \\
  = \bigl(x_i - y_i^- - \xi(y_i^-)\bigr)
  \int_{y_i^-}^{y_i}\!\! (u_0(y)t - \tilde\xi_t(y))\rho_0(y)\, dy\\
  + \bigl(x_i - y_i^+ - \xi(y_i^+)\bigr)
  \int_{y_i}^{y_i^+}\!\! (u_0(y)t - \tilde\xi_t(y))\rho_0(y)\, dy.
\end{multline}

Now take into account that at \(y_i^- \le z \le y_i^+\)
\begin{multline}
  \int_{y_i^-}^z (u_0(y)t - \tilde\xi_t(y))\rho_0(y)\, dy \\
  = \int_{m_i^-}^{m(z)}
  \left(\pd{\Phi_t}m - \pd{(\conv\Phi_t)}m\right)\, dm \\
  = \Phi_t(m(z)) - \conv\Phi_t(m(z)) \ge 0,
\end{multline}
and this integral vanishes only when \(z = y_i^+\). 
Therefore
\begin{multline}
  \int_{y_i}^{y_i^+}\!\! (u_0(y)t - \tilde\xi_t(y))\rho_0(y)\, dy \\
  = - \int_{y_i^-}^{y_i}\!\! (u_0(y)t - \tilde\xi_t(y))\rho_0(y)\, dy,
\end{multline}
and finally
\begin{multline}
  \int_{y_i^-}^{y_i^+} (x_i - y - \xi(y))
  (u_0(y)t - \tilde\xi_t(y))\rho_0(y)\, dy\\
  = (y_i^+ + \xi(y_i^+) - y_i^- - \xi(y_i^-)) \times \\
  \times (\Phi_t(m(y_i)) - \conv\Phi_t(m(y_i))) > 0.
\end{multline}
Thus the whole integral~\eqref{eq:77} is positive as a sum of positive
terms. Consequently, the displacement field \(\tilde\xi_t\) satisfies
the inequality~\eqref{eq:criterion} and minimizes the variational
principle \citepalias{S86}~\eqref{eq:6}.  Conversely, the unique
optimal displacement field in~\eqref{eq:6} must minimize the
variational principle~\citepalias{ERS96}.

\section{Ballistic aggregation in cylindrically and spherically
  symmetric cases}
\label{sec:spherical}

Let the initial density~\(\rho_0\) and velocity~\(\uu_0\) fields
depend on the \(d\)-dimensional vector~\(\yy\) and be symmetric with
respect to the origin:
\begin{equation*}
  \rho_0(\yy) = \rho_0(\abs\yy),\quad
  \uu_0(\yy) = \frac{u_0(\abs\yy)}{\abs\yy}\yy.
\end{equation*}
For \(d = 2\) this symmetry is cylindrical and for \(d = 3\),
spherical.  The results of this section hold for any \(d >
1\), including the nonphysical case of higher dimensions.

To extend the construction of the variational
principle~\citepalias{ERS96} to this case one has to replace the mass
element \(\rho_0(y)\,dy\) by \(\rho_0(\abs\yy)\Omega_d\, d\yy\) in
\eqref{eq:GPrinciple} and~\eqref{eq:cluster}.  Here \(\Omega_d =
2\pi^{d/2}/\Gamma(d/2)\) is the (hyper)surface area of the unit sphere
in \(d\)-dimensional space.  Arguing similarly to
Section~\ref{sec:clusters}, introducing the mass coordinate
\begin{equation*}
  m(\abs\yy) = \int_0^{\abs\yy}\rho_0(\eta)\Omega_d\, d\eta,
\end{equation*}
and setting
\begin{gather*}
    \Phi_0(m) = \int_0^{y(m)} \eta\rho_0(\eta)\Omega_d\, d\eta,\\
    U_0(m) = \int_0^{y(m)} u_0(\eta)\rho_0(\eta)\Omega_d\, d\eta,\\
    \Phi_t(m)=\Phi_0(m)+tU_0(m),\\
\end{gather*}
where
\begin{equation*}
  y(m) = \od{\Phi_0(m)}m,\quad u_0(y(m))=\od{U_0(m)}m,
\end{equation*}
we get an explicit solution that formally coincides
with~\eqref{eq:ERSvariational}, but with a new definition of the
function~\(\Phi_t\):
\begin{equation*}
  x(y(m), t) = \od{\bigl(\conv\Phi_t(m)\bigr)}m.
\end{equation*}

\section{On constructing solutions in the non-symmetric case}
\label{sec:counterexamples}

Can the above explicit construction of solutions to the ballistic
aggregation problem be extended to the case of non-symmetric flows?

An existence theorem for a solution to the ballistic aggregation
problem that respects the mass~\eqref{eq:mass} and
momentum~\eqref{eq:momentum} conservation laws was proved in
\citep{S01}.  However, this proof is non-constructive and thus gives
little information about the structure of the solution and provides no
method for approximating it numerically.

On the contrary, both variational principles \citepalias{S86} and
\citepalias{ERS96} provide structural information and approximation
procedures.  Moreover, for a potential initial velocity field
\(\uu_0(\yy)\) these variational principles have natural
multidimensional extensions requiring no specific assumptions on
symmetry of the flow.  However it is possible to construct examples of
asymmetric flows for which these extensions give physically incorrect
answers, as we show below.

\subsection{A multidimensional extension of the variational principle
  \citepalias{ERS96}}
\label{sec:sinaicounter}

To extend the variational principle
\citepalias{ERS96}~\eqref{eq:ERSvariational} to the multidimensional
case, introduce a vector ``mass'' coordinate~\(\mm\) of the same
dimension as the vector~\(\yy\),
\begin{equation*}
  \abs{\pd\mm\yy} = \rho_0(\yy),
\end{equation*}
and define functions \(\Phi_0(\mm)\), \(U_0(\mm)\) such that
analogues of equations~\eqref{eq:12} hold:
\begin{equation*}
  \yy(\mm) = \nabla_\mm\Phi_0,\quad \uu_0(\mm) = \nabla_\mm U_0.
\end{equation*}
Then it is natural to define \(\xx(\yy, t)\) similarly
to~\eqref{eq:ERSvariational}:
\begin{equation}
  \label{eq:14}
  \xx(\yy(\mm), t) = \nabla_\mm \conv(\Phi_t + U_0 t).
\end{equation}

Whatever the way in which \(\mm(\yy)\), \(\Phi_0(\mm)\)
and~\(U_0(\mm)\) are defined in general, for the uniform initial
density \(\rho_0(\yy) \equiv 1\) it suffices to set
\begin{equation}
  \label{eq:15}
  \mm = \yy,\quad \Phi_0(\mm) = \frac{{\abs\mm}^2}2
\end{equation}
and take \(U_0\) for the initial velocity field potential~\(\uu_0\).
Still even in this case it is possible to show that the proposed
modification of the variational principle~\citepalias{ERS96} leads to
incorrect results.

Consider a spherical wave expanding fom the origin.  Suppose that
initially matter is at rest while the velocity potential~\(U_0\) is
constant everywhere except a small area about the origin.
Asymptotically for large times the size of this area and details of
the velocity field inside it become inessential and the initial
velocity potential can be approximated with
\begin{equation*}
  U_0(\yy) =
  \begin{cases}
    0, & \yy = 0,\\
    U > 0, & \abs\yy > 0.
  \end{cases}
\end{equation*}
At time~\(t>0\) the radius~\(R(t)\) of the spherical wave is defined
by the derivative \(\od{(\conv\Phi_t)}{\abs\yy}\) at \(\abs\yy = 0\),
where
\begin{multline}
  \label{eq:17}
  \Phi_t(\yy) = \frac{{\abs\yy}^2}2 + U_0(\yy)t,\\
  \conv\Phi_t(\yy) =
  \begin{cases}
    \sqrt{2Ut}\abs\yy, & 0\le\abs\yy\le\sqrt{2Ut},\\
    \frac{{\abs\yy}^2}2 + Ut, &\sqrt{2Ut}\le\abs\yy.
  \end{cases}
\end{multline}
So the proposed extension of the variational
principle~\citepalias{ERS96} predicts that the radius of the spherical
wave grows as~\(\sqrt{2Ut}\) regardless of the dimension of space.

On the other hand, the full mass of the wave front at time~\(t\) is
equal to the mass of matter initially distibuted over the area
contained within the wave front, i.e., proportional to~\(R^d\), where
\(d = 2\) in the cylindrical case and \(d = 3\) in spherical case.
Since matter moves only in radial directions, the full momentum is
preserved in any cone with the vertex at the origin.  Hence the
product of the velocity of the wave front by its mass must stay
constant: \(R^d\od Rt = \text{const}\), so that
\begin{equation}
  \label{eq:33}
  R(t) \propto t^{1/(d + 1)}.
\end{equation}
It is this time dependence of~\(R\) that appears in
Section~\ref{sec:spherical}, while \eqref{eq:17} and~\eqref{eq:14}
turn out to hold only for~\(d = 1\).

It is interesting to determine the time dependence of the kinetic
energy~\(E(t)\).  The full kinetic energy of the flow is the product
of the mass of the vawe front by its squared radial
velocity. Accordingly, in the physically correct
solution~\eqref{eq:33} we have \(E \propto R^{-d} \propto t^{-d/(d +
  1)}\), while formula~\eqref{eq:14} gives \(E \propto t^{(d - 2)/2}\);
for \(d = 3\), the kinetic energy of the spherical wave turns out to
grow indefinitely!

Although the choice of mass coordinates~\(\mm(\yy)\) is not uniquely
determined by~\(\rho_0\) alone, this freedom is not enough to resolve
this contradiction.  For a constant initial density~\(\rho_0\) one can
show~\citep{CL01} that the only way to introduce mass coordinates
while preserving the spherical symmetry is~\eqref{eq:15}.  Thus the
convex hull construction of the variational
principle~\citepalias{ERS96} cannot be extended to the non-symmetric
multidimensional case.

Another analysis showing incorrectness of this extension of the
variational principle~\citepalias{ERS96} is given in~\citep{R02}.

\subsection{A multidimensional extension of the variational principle
  \citepalias{S86}}
\label{sec:shnirelmancounter}

To extend the variational principle \citepalias{S86} to the
multidimensional case, one has to define a set of feasible
displacement fields in such a way that a solution to the ballistic
aggregation problem would still be defined by an orthogonal projection
construction.

In one dimension a displacement field~\(\xi(y)\) is feasible when the
function \(x(y) = y + \xi(y)\) is monotonic, or equivalently when its
primitive function is convex.  Observe that, regardless of the
dimension, the `mixture' of two convex functions \(\Phi\),~\(\Psi\) of
the form \((1 - \alpha)\Phi + \alpha\Psi\), where \(0 \le \alpha \le
1\), is always convex and that the differentiation operation is
linear.  Therefore the class of gradients of convex functions in any
dimension is a convex subset of a suitable functional space.  Call the
displacement field \(\xixi(\yy) = \xx(\yy) - \yy\) \emph{feasible} if
\(\xx(\yy)\) is the gradient of a convex function.  Then a natural
multidimensional extension of the variational
principle~\citepalias{S86} consists in minimizing the norm of
discrepancy
\begin{equation}
  \label{eq:20}
  \int \abs{\xixi_t(\yy) - \uu_0(\yy)t}^2\rho_0(\yy)\, d\yy = \min
\end{equation}
over the class of feasible displacement fields defined as above.

Suppose that the initial velosity field~\(\uu_0\) is potential, so that
\begin{equation*}
  \uu_0(\yy) = \nabla U_0(\yy),\
  \xx(\yy, t) = \yy + \nabla U_0(\yy)t = \nabla\Phi_t(\yy),
\end{equation*}
where
\begin{equation*}
  \Phi_t(\yy) = \frac12{\abs\yy}^2 + U_0(\yy)t.
\end{equation*}
Evidently, at \(t = 0\) the function~\(\Phi_0(\yy) = {\abs\yy}^2/2\)
is convex.

As long as the Jacobian of the map \(\yy \mapsto \xx(\yy, t)\) is
non-zero (positive), trajectories of particles do not cross.  Since
this Jacobian coincides with the determinant of the matrix of second
derivatives (the Hessian matrix) \((\partial_i\partial_j \Phi_t)\),
convexity of the function~\(\Phi_t(\yy)\) is preserved.  Therefore the
displacement field \(\uu_0(\yy) t\) stays feasible and the discrepancy
norm~\eqref{eq:20} is equal to zero, so the proposed form of the
variational principle \citepalias{S86} gives a correct solution in
this case.

After crossing of two infinitesimally close trajectories the
determinant of the Hessian matrix vanishes, the function
\(\Phi_t(\yy)\) ceases to be convex, and the displacement field
minimizing the~\eqref{eq:20} can no longer be given by~\(\uu_0 t\).
Similarly to the argument of Section~\ref{sec:shnirelman} one can show
that in the spherically symmetric case the proposed form of the
variational principle \citepalias{S86}~\eqref{eq:20} gives a correct
solution.  However the following example shows that the proposed
extension gives wrong results in less symmetrical cases.

\begin{figure}
  \centering\includegraphics{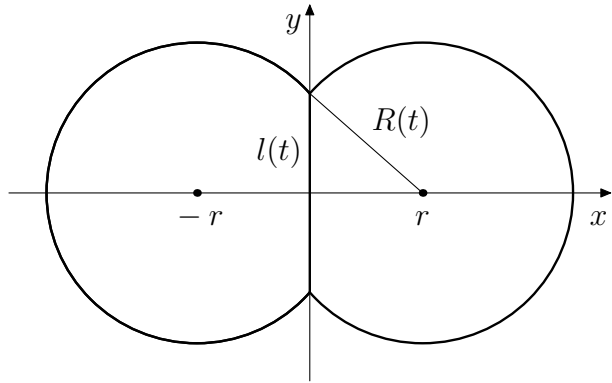}
  \caption{Two cylindrical waves: an example of a flow for which the
    proposed multidimensional extension of the variational principle
    \citepalias{S86}~\protect\eqref{eq:20} gives an incorrect
    result. The front is shown at time~\(t\) satisfying
    \((r/K)^3<t<t^* = (5/2)^{5/2}(r/K)^3\).}
  \label{fig:2spheres}
\end{figure}

Consider two identical cylindrical waves in a two-dimensional space
with centers separated from each other by~\(2r\) and radii growing as
\(R(t) = Kt^{1/3}\) in accordance with~\eqref{eq:33}.  At time \(t >
(r/K)^3\) the fronts of these waves start to coalesce, giving rise to
a flat portion of the wave front with matter moving longitudinally
(Fig.~\ref{fig:2spheres}).  We now construct the displacement field
describing the distribution of matter in the cylindrical portions and
in the flat portion of the wave front for \(t > (r/K)^3\).

Let the coordinates be chosen so that centers of waves are at \((\pm
r, 0)\).  At time~\(t\) the flat portion is the segment of length
\(2l(t) = 2\sqrt{R^2(t) - r^2}\) in the \(y\)~axis.  Matter coming to
this segment from the rounded portions of the wave front sticks
together and moves outward with velocity \(v = \dot R\cdot l/R\).
Observe that the length of the segment increases faster, at the rate
\begin{equation*}
  \dot l = \frac{R\dot R}{\sqrt{R^2 - r^2}} = \dot R\cdot \frac Rl
  > \dot R\cdot \frac lR = v;
\end{equation*}
thus endpoints of the flat segment do not affect the motion of matter
inside it.

Let \(T(l)\) be the inverse funcion for~\(l(t)\),
\begin{equation*}
  T(l) = \frac 1{K^3} (l^2 + r^2)^{3/2},
\end{equation*}
and take~\(l\) as a parameter.  Since \(R(t) = Kt^{1/3}\), we have
\(\dot R/R = 1/3t\), so that \(v(l) = l\dot R/R = l/3T(l)\).
Therefore at time \(t > T(l)\) the particle parametrized with~\(l\)
will be located at
\begin{equation*}
  y(l, t) = l + v(l)\bigl(t - T(l)\bigr) = \frac 23l + \frac {tl}{3T(l)}.
\end{equation*}
The distribution of mass inside the flat segment stays continuous as
long as the map \(l\mapsto y(l, t)\) is monotonic, i.e., while \(\pd
yl = 2/3 + (t/3)(l/T(l))' > 0\) for all~\(l\).  This inequality is
satisfied for \((r/K)^3 < t < t^*\), where
\begin{equation*}
  t^* = \min_{l>r/\sqrt2}\left(-\frac 2{(l/T(l))'}\right)
  = \left(\frac52\right)^{5/2}\left(\frac rK\right)^3.
\end{equation*}
Here minimization is performed for~\(l>r/\sqrt2\) because for \(0\le
l\le r/\sqrt2\) the condition \(\pd yl\ge 0\) holds for any~\(t\).
\begin{figure}[h]
  \centering\includegraphics{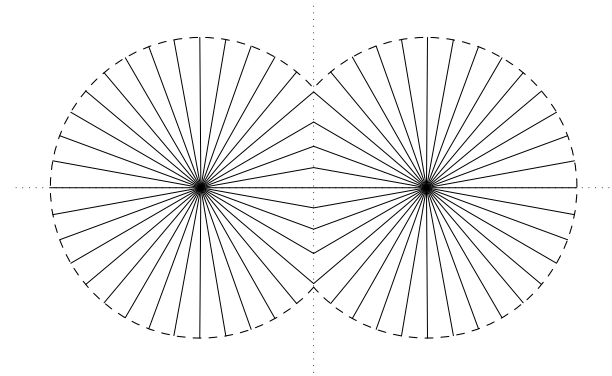}
  \caption{Continuous black lines starting at \((\pm r, 0)\) show
    initial positions of particles located at the points of the wave
    front in the flow of Fig.~\protect\ref{fig:2spheres} (for times
    \((r/K)^3 < t < t^* = (5/2)^{5/2}(r/K)^3\)).  Note that particles
    forming the flat portion of the front are initially located on
    two-segment broken lines in the central area of the figure.}
  \label{fig:lagrange2spheres}
\end{figure}

Observe now that initial positions of particles that are located at
the point \((0, y(l, t))\) in the flat segment of the front at time
\(T(l)\le t<t^*\) form a two-segment broken line connecting the point
\((0, l)\) with centers of both waves
(Fig.~\ref{fig:lagrange2spheres}).  On the other hand, the preimage of
any point in a mapping defined by the gradient of a convex function
must itself be a convex set (a point, a segment, or a convex domain).
Since preimages of some points in the displacement field are
non-convex broken lines, this displacement field cannot be realized by
the gradient of a convex function, i.e., it is not `feasible' in the
above sense and the proposed natural extension of the variational
principle \citepalias{S86} provides an incorrect description of
describes the motion of matter in the case of two cylindrical waves.


\end{document}